%%%%%%%%%%%%%%%%%%%%%%%%%%%%%%%%%%%%%%%%%%%%%%%%%%%%%%%%%%%%%%%%%%%%%%
% Filename: stgr9220.tex from stgr1030.tex
% Title   : Symmetric teleparallel general relativity                %
% Authors : James M. Nester and Hwei-Jang Yo                         %
% Date    : 20 Feb 1999                                              %
% filetype: TeX                                                      %
%%%%%%%%%%%%%%%%%%%%%%%%%%%%%%%%%%%%%%%%%%%%%%%%%%%%%%%%%%%%%%%%%%%%%%

\magnification=\magstep1
\hoffset -7mm
%--------------------------
% FONTS
\font\seventeenrm=cmr17
%--------------------------
%\pageno=0
%\baselineskip=24 true pt plus1pt minus1pt
\footline={\hfil\tenrm\folio\hfil}

%\hsize=16 true cm

\line{\hfil NCU-CCS-980904}
\line{\hfil /xxx.lanl.gov/gr-qc/9809049}
\bigskip\bigskip
\centerline{\seventeenrm
Symmetric teleparallel general relativity}
\bigskip

\centerline{James M. Nester\footnote{*}{email: nester@joule.phy.ncu.edu.tw}
 and Hwei-Jang Yo\footnote{**}{email: s2234003@twncu865.ncu.edu.tw}}
\medskip
\centerline{\it Department of Physics and Center for Complex Systems,}
\centerline{\it National Central University, Chungli, Taiwan 320, R.O.C.}
 \bigskip
\bigskip



\line{\hfil $\qquad$ (fundamental problems and general formalism)\hfil}

\bigskip\bigskip
%\centerline{\underbar{Summary}}
\centerline{ABSTRACT}
\midinsert
\narrower\narrower

General relativity can be presented in terms of other geometries
besides Riemannian.  In particular, teleparallel geometry (i.e.,
curvature vanishes) has some advantages, especially concerning
energy-momentum localization and its ``translational gauge theory''
nature.  The standard version is metric compatible, with torsion
representing the gravitational ``force''.  However there are many
other possibilities.  Here we focus on an interesting alternate
extreme:  curvature and torsion vanish but the nonmetricity $\nabla g$
does not---it carries the ``gravitational force''.  This {\it
symmetric teleparallel} representation of general relativity
covariantizes (and hence legitimizes) the usual coordinate
calculations.  The associated energy-momentum density is essentially the
Einstein pseudotensor, but in this novel geometric representation it
is a true tensor.

\endinsert
\vfil\eject
%\smallskip

\vfil\eject

Alternate representations of a theory often provide valuable insight.
For Einstein's gravitational theory, general relativity (GR), there
are alternate variables [1, 2, 3] and even alternate geometries [4].
The standard representation of GR has a Riemannian geometry with the
curvature representing the ``gravitational tidal force''.  In
contrast, for the alternate geometries considered here, the curvature
{\it vanishes}.

Such {\it teleparallel} (a.k.a.\ absolute parallel, fernparallelismus,
Weitzenb\"ock) geometries [5, 6] have a venerable history.  Einstein
himself used such a geometry in one of his unified field theory
attempts [7].  Others have used it to formulate alternate gravity
theories, e.g., [8].  Of interest here, however, is that GR can
be recast in terms of teleparallel geometry, see e.g., [9, 10].  Of late, such
a formulation has attracted renewed attention including [2, 11, 12,  13,
14, 15].  A major advantage concerns energy-momentum, its representation,
positivity and localization [9, 11, 13].  Another appeal is that the
teleparallel formulation can be regarded as a ``translational gauge
theory'' [10, 13, 14, 15].

For a teleparallel geometry there is a preferred class of frames,
which greatly simplify computations.  They can be obtained by
selecting any frame at one point and parallel transporting it to all
other points.  Since the curvature vanishes, the parallel transport is
path independent so the resultant frame field is globally well
defined---if the manifold is parallelizable (a strong global
topological restriction but highly desirable physically---being
necessary for spinor fields and the Cauchy problem).  In such a
teleparallel frame field the connection coefficients vanish.
Teleparallel frame fields are unique up to a global (constant, rigid)
linear transformation.  Parallel transport in a teleparallel frame
field is simply accomplished by keeping the coefficients of tensors
constant, and the covariant derivative simply reduces to the partial
derivative.

The standard teleparallel representation of GR has a metric compatible
connection.  While the curvature vanishes the torsion does not; it
acts like a gravitational ``force''.  The simplest description is
in terms of orthonormal-teleparallel frames.  This representation is
easily constructed:  given the metric for GR, simply choose {\it any}
orthonormal frame and declare it to be parallel (i.e., introduce a new
connection, the one which has vanishing coefficients in this
particular frame).  Thus each Riemannian geometry is represented by
not just one but rather a whole gauge equivalence class of
teleparallel geometries!  (The relevant geometric point has been
nicely stated recently:  ``Strictly speaking, there is no such thing
as curvature or torsion of spacetime, but only curvature and torsion
of connections'' [14].)

Such a teleparallel geometry was used by M\o{}ller [9] to define a
gravitational energy-momentum tensor (not a pseudotensor).  {\it
Geometrically} it is a tensor---but the catch is that the whole
geometrization is rotationally gauge dependent.  This seeming
liability of the orthonormal-teleparallel representation of GR has
more recently proved advantageous:  it has been exploited, via a
suitable rotational gauge condition [16], to obtain a Hamiltonian
based tensorial (not spinorial) positive energy proof [11].

The orthonormal-teleparallel representation of Einstein's theory is
indeed interesting and useful (it has even been referred to as the
{\it teleparallel equivalent of general relativity} [13]).  However
there are other possibilities which, as far as we can ascertain, have
been generally overlooked (aside from a few hints like [17] \S5.9 and
[18] p32).

One could consider a general---non-metric compatible,
non-symmetric---teleparallel connection (this is a special case of the
Metric Affine type of geometry, see [17] \S 5.9 and [6] p142).  Each
possibility can be simply represented by a choice of frame and
vanishing connection coefficients.  Thus general relativity has a
great many teleparallel representations.  There may even be some nice
choices which have the torsion carrying part of the gravitational
force and the non-metricity the rest.  However, here we focus on just
one interesting extreme:  the {\it symmetric teleparallel} formulation
of general relativity (STGR) in which the torsion vanishes but the
nonmetricity $\nabla g$ does not (this is just the opposite of the
usual choice).

Geometry is determined by the metric and the connection.
The latter is conveniently represented
by the one-form coefficients determined by the covariant differential
$\nabla e_\beta=e_\alpha\Gamma^\alpha{}_\beta$,
where $e_\alpha$ is a general frame field with dual
 coframe $\vartheta^\alpha$.  They determine the
{\it curvature}
$R^\alpha{}_\beta:=d\Gamma^\alpha{}_\beta+\Gamma^\alpha{}_\gamma
\wedge\Gamma^\gamma{}_\beta$
 and {\it torsion}
$T^\alpha={1\over2}\,T^\alpha{}_{\mu\nu}\vartheta^\mu\wedge\vartheta^\nu
:=d\vartheta^\alpha + \Gamma^ \alpha{}_\gamma\wedge\vartheta^\gamma$
two-forms.  The connection can be decomposed into the
Levi-Civita connection and deformation one-forms:
$$\Gamma^\alpha{}_\beta=
\Gamma^{\{\}}{}^\alpha{}_\beta-A^\alpha{}_\beta.\eqno(1)$$
The general decomposition of the deformation ([6] p141, [17] \S3.10),
$$A_{\alpha\beta}=K_{\alpha\beta}-{1\over2}Q_{\alpha\beta}
-Q_{\gamma[\alpha\beta]}\vartheta^\gamma, \eqno(2)$$
includes the contortion,
$K_{\alpha\beta}=K_{\alpha\beta\gamma}\vartheta^\gamma$,
which is linear in the torsion,
$$K_{\alpha\beta\gamma}={1\over2}(T_{\beta\alpha\gamma}-
T_{\alpha\beta\gamma}+T_{\gamma\alpha\beta}),\eqno(3)$$
and nonmetricity
$Q_{\alpha\beta}=Q_{\alpha\beta\gamma}\vartheta^\gamma
=-Dg_{\alpha\beta}$.
It induces an associated decomposition of the curvature:
$$R^\alpha{}_\beta\equiv R^{\{\}}{}^\alpha{}_\beta-D^{\{\}}A^\alpha{}
_ \beta+A^\alpha{}_\gamma\wedge A^\gamma{}_\beta.\eqno(4)$$

An advantage of the teleparallel formulation appears in the
variational principle.  The standard Einstein-Hilbert scalar curvature
Lagrangian density is asymptotically $O(1/r^3)$, so the action
diverges.  This can be improved to a convergent $O(1/r^4)$ by removing
a total derivative (leaving the field equations unchanged) but the
resulting density is not covariant --- the Noether arguments then give
energy-momentum {\it pseudo\/}tensors.  In contrast, for the teleparallel
formulations, removal of the total derivative leaves a covariant,
asymptotically convergent, Lagrangian which generates a covariant
energy-momentum tensor.

It is convenient to introduce the dual basis for forms:
$\eta^{\alpha\beta\cdots}:=
*(\vartheta^\alpha\wedge\vartheta^\beta\cdots)$,
in terms of which the scalar curvature decomposes as
$$\eqalignno{
R\eta=R^\alpha{}_\beta\wedge\eta^\beta{}_\alpha
&=R^{\{\}}{}^\alpha{}_\beta\wedge\eta^\beta{}_\alpha
-D^{\{\}}A^\alpha{}_\beta\wedge\eta^\beta{}_\alpha
+A^\alpha{}_\gamma\wedge A^\gamma{}_\beta\wedge\eta^\beta{}_\alpha\cr
&=R^{\{\}}\eta-d(A^\alpha{}_\beta\wedge\eta^\beta{}_\alpha)
-A^\alpha{}_\beta\wedge
D^{\{\}}\eta^\beta{}_\alpha+A^\alpha{}_\gamma\wedge
A^\gamma{}_\beta\wedge\eta^\beta{}_\alpha. &(5)\cr}$$
But
$D^{\{\}}\eta_\alpha{}^\beta$
vanishes (because the torsion and
nonmetricity vanish for the Levi-Civita connection); hence, with
$\Gamma^\alpha{}_\beta$
teleparallel, removing an exact differential from the Einstein-Hilbert
scalar curvature Lagrangian 4-form, gives the (note: it's covariant)
general teleparallel Lagrangian 4-form:
$${\cal L}=A^\alpha{}_\gamma \wedge
A^\gamma{}_\beta\wedge\eta^\beta{}_\alpha,\eqno(6)$$
to which one can adjoin the vanishing curvature condition via a
Lagrange multiplier.

The special case $T\ne0=Q$ has been studied extensively, e.g., [9, 10, 11,
13, 14, 15].  Here we examine the opposite case:  $T=0\ne Q$.  Then
${\cal L}$ becomes quadratic in $Q$.  More generally there are 5
independent quadratic Q terms [19, 20],
and hence a whole 5-parameter class of symmetric teleparallel
theories that really merit investigation.  We expect that most would
have new types of ``gravitational'' effects.  A general investigation
is underway; only the one special case equivalent to GR is considered
here.

While the symmetry condition can be introduced into the action via
a Lagrange multiplier,
the simpler representation is in terms of a teleparallel frame where
$\Gamma$ vanishes.  The symmetry condition then reduces to
$T^\alpha=d\vartheta^\alpha=0$.
Consequently (at least locally)
$\vartheta^\alpha=dx^\alpha$
for some coordinate system and covariant derivatives reduce to coordinate
partial derivatives.  The non-metricity tensor simplifies to
$Q_{\mu\nu\lambda}=- g_{\mu\nu,\lambda}$,
and  the deformation tensor acquires the Christoffel symbol form:
$$
A^\alpha{}_{\beta\gamma}=
{1\over2}g^{\alpha\lambda}(
Q_{\beta\lambda\gamma}+Q_{\gamma\lambda\beta}-Q_{\beta\gamma\lambda})=
-{1\over2}g^{\alpha\lambda}(
\partial_\gamma
g_{\beta\lambda}+\partial_\beta
g_{\gamma\lambda}-\partial_\lambda g_{\beta\gamma}
)=-\Bigl\{ {\alpha \atop \beta\gamma} \Bigr\},
\eqno(7)$$
as expected from Eq.\ (1).
Consequently the associated Lagrangian density becomes
 $${\cal L}=\sqrt{-g}g^{\mu\nu}\biggl(
\Bigl\{ {\alpha \atop \gamma\mu} \Bigr\}
\Bigl\{ {\gamma \atop \nu\alpha} \Bigr\} -
\Bigl\{ {\alpha \atop \gamma\alpha} \Bigr\}
\Bigl\{ {\gamma \atop \mu\nu} \Bigr\}
\biggr),\eqno(8)$$
which has often been used in the past.  Notwithstanding appearances,
we want to emphasize that, in terms of the STGR geometry this
Lagrangian---and indeed a host of apparently non-covariant
calculations done in numerous works especially in the early days---are
viewed as covariant.  Consider just one example:  Tolman's calculation
of the energy-momentum density for gravity [21].  The result, the
Einstein pseudotensor, from the STGR viewpoint is a covariant object.
The coordinate dependence of the energy-momentum density is now
elevated from a choice of reference frame to a ``gauge'' choice of
geometry.  Thus the associated energy-momentum density is a covariant
(but geometry gauge dependent) tensor.

This novel representation, in a kind of {\it tour de force}, transforms
all the usual coordinate calculations, rendering them into a
geometrically covariant form and thereby bestowing upon them a greater
degree of legitimacy.  In our opinion, the result is not merely
cosmetic.  Transforming the usual coordinate invariance into a gauge
invariance associated with a connection of essentially the same type
as used in other gauge theories invites a new teleparallel approach to
gravity as a gauge theory---one based on a symmetric connection and
nonmetricity rather than on metric compatibility and torsion.
The STGR formulation brings a new perspective to bear on GR.  We don't
yet know what might be revealed.  Certainly the STGR representation
offers a convenient way to formally treat gravity like other fields.
The gravitational interaction effects, via the nonmetricity, have a
character much like a Newtonian force and are derived from a
potential, the metric; nevertheless the formulation is geometric and
covariant.

Of course the STGR formulation has some liabilities.  It must be
emphasized that in this geometry it is no longer possible to simply
commute derivatives and the raising or lowering of indices via the
metric as we are so accustomed to do in the standard Riemannian
approach.  Hence tensorial equations will appear differently depending
on how the indices are arranged.  However this apparent computational
complication is to a large extent offset by the fact that in the
teleparallel geometry covariant derivatives commute (since curvature
vanishes).  It should also be noted that the coupling to sources which
effectively reproduces the standard GR interaction is {\it not}
teleparallel minimal coupling.  (This serves to draw attention to the
fact that minimal coupling is not a sacred principle.)\ \ Another
obvious limitation of the STGR formulation is that it (almost)
requires a global coordinate system.

Of course many investigators have treated gravity as a field theory in
flat spacetime without developing a teleparallel geometric formulation
(see, e.g., Ch VI.\ \S5 in [22] and Ch.\ 11  in Thorne's
popular book [23]; the latter includes an interesting discussion of the
philosophical issues).  \ \ The STGR alternative offers our old familiar
theory a new symmetric teleparallel apparel in which it acquires a
different geometric appearance.

\beginsection ACKNOWLEDGMENTS

The authors appreciated comments by Kip Thorne and F.~W.~Hehl as well as
a suggestion of the referee.
This work was supported by the National Science Council of the
R.O.C. under contracts NSC87-2112-M-008-007, NSC88-2112-M-008-018.

\vfil\eject
\beginsection REFERENCES

\frenchspacing

\item{[1]} A. Ashtekar,  {Phys. Rev. D \bf 36}, 1587 (1987).

\item{[2]} E. W. Mielke,  {Ann. Phys. \bf 219}, 78--108 (1992).

\item{[3]} J. M. Nester and R. S. Tung,  {Gen. Rel. Grav.
\bf 27}, 115-119 (1995), (Fourth award, Gravity Research Foundation 1994).

\item{[4]} M. Ferraris and J. Kijowski,
  {Gen. Rel. Grav. \bf 14}, 165--180 (1982).

\item{[5]} R. Weitzenb\"ock,  {\it Invariantentheorie}
(Noordhoff, Gronningen, 1923).

\item{[6]} J. A. Schouten,  {\it Ricci Calculus}, 2nd
edition (Springer-Verlag, London, 1954).

\item{[7]} A. Einstein,  {Sitzungsber. Preuss. Akad. Wiss.}  217 (1928).

\item{[8]} K. Hayashi and T. Shirafuji,
 {Phys. Rev. D \bf 19}, 3524--3553 (1979).

\item{[9]} C. M{\o}ller,
  {Mat. Fys. Skr. Dan. Vid. Selsk. \bf 1}, 1-50 (1961).

\item{[10]} Y. M. Cho,  {Phys. Rev. D \bf 14}, 2521--2525 (1976).

\item{[11]} J. M. Nester,
 {Int. J. Mod. Phys. A \bf 4}, 1755--1772 (1989).

\item{[12]} R. P. Wallner,  {Phys. Rev. D \bf 42}, 441--448 (1990).

\item{[13]} J. W. Maluf,  {J. Math. Phys. \bf 35}, 335--343 (1994);
  {J. Math. Phys. \bf 36}, 4242--4247 (1995);
 {Gen. Rel. Grav. \bf 28}, 1361--1376 (1996).

\item{[14]} V. C. Andrade and J. G. Pereira,
 {Phys. Rev. D \bf56}, 4689--4695 (1997);
 {Gen. Rel. Grav. \bf 30}, 263 (1998).

\item{[15]} U. Muench, F. Gronwald and F. W. Hehl,
{Gen. Rel. Grav. \bf 30}, 933--961 (1998).

\item{[16]} J. M. Nester,  {J. Math. Phys. \bf 33}, 910--913 (1992).

\item{[17]} F. W. Hehl, J. D. McCrea, E. W. Mielke and Y. Ne'eman,
 {Phys. Rep. \bf 258}, 1--171 (1995).

\item{[18]} R. M. Wald,  {\it General Relativity}
 (Univ. of Chicago, Chicago, 1984).

\item{[19]} Y.-S. Duan, J.-C. Liu and X.-G. Dong,
{Gen. Rel. Grav. \bf 20}, 485--496 (1988).

\item{[20]} Yu.N. Obukhov, E. J. Vlachynsky, W. Esser, and F. W. Hehl,
{Phys. Rev. D \bf 56}, 7769--7778 (1997).

\item{[21]} R. C. Tolman  {Phys. Rev. \bf 35}, 875--881 (1930).

\item{[22]} J. L. Synge, {\it Relativity: the General Theory},
(North-Holland, Amsterdam, 1960).

\item{[23]} K. S. Thorne, {\it Black Holes and Time Warps: Einstein's
Outrageous Legacy}, (Norton, New York, 1994).

 \bye